## Article

# A Rational Design Method for the Nagoya Type-III Antenna

Daniele Iannarelli [1], Francesco Napoli [2], Antonella Ingenito [1,*], Alessandro Cardinali [3,4], Antonella De Ninno [2] and Simone Mannori [2]

[1] School of Aerospace Engineering, Sapienza University of Rome, Via Salaria 851-881, 00138 Rome, Italy; daniele.iannarelli@uniroma1.it
[2] ENEA Frascati Research Center, Via Enrico Fermi 45, 00044 Frascati, Italy; francesco.napoli@enea.it (F.N.)
[3] Istituto Nazionale di Astrofisica (INAF)—Istituto di Astrofisica e Planetologia Spaziali, Via del Fosso del Cavaliere 100, 00133 Roma, Italy; alessandro.cardinali@sapienza.isc.cnr.it
[4] Consiglio Nazionale delle Ricerche (CNR), Istituto Sistemi Complessi, Politecnico di Torino, Corso Duca degli Abruzzi 24, 10129 Torino, Italy
* Correspondence: antonella.ingenito@uniroma1.it

**Abstract:** The current study, as part of a PhD project on the design of a helicon thruster, aims to provide a rational methodology for the design of the helicon thruster's main component, i.e., the helicon antenna. A helicon thruster is an innovative electrodeless plasma thruster that works by exciting helicon waves in a magnetized plasma, and its antenna is capable of producing a uniform, low-temperature, high-density plasma. A magnetic nozzle is used to accelerate the exhaust plasma in order to generate a propulsive thrust. In this paper, we consider a simple helicon antenna, specifically the Nagoya type-III antenna. We consider a common experimental setup consisting of a quartz tube with finite length containing a uniform magnetized plasma and a Nagoya type-III antenna placed at the tube centre. Considering previous studies on helicon waves theory, we compare three different design methods, each based on simplifying different modelling assumptions, and evaluate the predictions of these models with results from full-wave 3D simulations. In particular, we concentrate on deriving a rational design method for the helicon antenna length, given the dimension of the quartz tube and the desired target plasma parameters. This work aims to provide a practical and fast method for dimensioning the antenna length, useful for initializing more accurate but computationally heavier full-wave simulations in 3D geometry or simply for a rapid prototyping of the helicon antenna. These results can be useful for the development of a helicon thruster but also for the design of a high-density radiofrequency plasma source.

## 1. Introduction

This work is part of a PhD research project, the result of collaboration between the School of Aerospace Engineering (SIA) of La Sapienza University in Rome and ENEA Frascati Research Centre, aimed at the design and development of an RF plasma source for a helicon thruster. A helicon thruster [1,2] is a relatively new electrodeless plasma thruster that works by exciting helicon waves in a magnetized plasma. The helicon thruster's main component is the helicon antenna, which is capable of producing a uniform, low-temperature (few eV), high-density plasma (electron density up to $10^{20}$ particles/m$^3$) [3]. In addition, a magnetic nozzle [4] produces the propulsive thrust accelerating the plasma generated from the helicon antenna. Helicon waves [5–12] belong to the category of whistler waves, which are a type of low-frequency circularly polarized electromagnetic waves propagating in magnetized plasmas, often found in Earth's ionosphere and magnetosphere. The dispersion of these waves in the magnetized plasma causes higher frequencies to travel faster than lower frequencies, leading to the characteristic "whistling" sound when detected on the ground or by instruments in space. Helicon waves differ from them in that they have a much lower frequency than the electron cyclotron frequency and that they are hybrid modes of bounded systems in which their purely electromagnetic character is lost [7]. Helicon waves have attracted great interest since 1970 when Boswell [5,6] discovered that dense plasmas could be generated by exciting helicon waves with a simple antenna in a magnetized plasma. However, it was soon realized that the collisional absorption rate of helicon waves predicted for those experiments was too low to account for the ionization efficiency of these waves [6]. In 1991, Chen [7] suggested that Landau damping could be the cause of the efficient energy absorption. This explanation was later used by many authors, however, Chen and Blackwell [8] experimentally demonstrated that the number of fast electrons was too small to be able to explain the high ionization efficiency, thus downgrading the role of Landau damping. Subsequently, research on helicon plasmas revealed other mechanisms that could explain their efficient plasma production [8–12]. In recent years, one of the more widely accepted explanations is the coupling of helicon waves with quasi-electrostatic waves called Trievelpiece–Gould (TG) waves [13]. Helicon waves can penetrate into the core region of the plasma where they can deposit a small fraction of their energy due to their low collisional damping rate but, by coupling with TG waves [9,10] which are excited mainly near the antenna and are strongly damped by collisions, a large fraction of their energy can be deposited at the plasma edge leading to a highly efficient production of plasma. Mode conversion with TG waves can occur at the plasma edge for uniform plasmas or in the presence of a density gradient for non-uniform plasmas [9,10,14]. The existence of TG waves has also been experimentally verified [15] and a general theory of helicon waves has been developed by Chen and Arnush [9,10]. Within this theoretical framework, taking into account the electron inertia, it is possible to retain the mode coupling with the TG waves and explain some distinct features observed in helicon discharges. However, the physics of helicon waves is still the subject of intense theoretical and experimental study to explain the rich phenomenology of helicon discharges, since some non-linear effects such as parametric instabilities and drift-wave instabilities [11,12] must be taken into account to explain anomalous power absorption and ion heating. In this paper, we consider a simple helicon antenna, specifically the Nagoya type-III antenna [7]. We consider a common experimental setup consisting of a quartz tube with given radius and finite length containing a uniform magnetized plasma and a Nagoya type-III antenna placed at the centre of the tube. The length of the quartz tube has been chosen to be much greater than the helicon wavelength in the axial direction to minimize the effects of end reflections. Taking into account previous studies on helicon wave theory [7,9,10,12], we compare three different design methods, each based on different simplifying modelling assumptions, and evaluate the predictions of these models with results from full-wave 3D simulations. In particular, we focus on deriving a rational design methodology for the helicon antenna length, given the quartz tube dimensions and desired target plasma parameters. The main objective of this work is to provide a practical and fast method for dimensioning the antenna length, useful for initializing more accurate but computationally expensive full-wave simulations in 3D geometry, or simply for rapid prototyping of the helicon antenna. These results may be useful for the development of a helicon thruster for space applications [1,2,4], but also for the design of a high-density RF plasma source for industrial applications [12].

## 2. Helicon Wave Modelling and Design Methods

Helicon waves can be excited in magnetized plasma by means of specific helicon antennas which are designed to resonate with the normal modes of magnetized plasmas, enabling the propagation of helicon waves. There are several types of helicon antennas that differ in their geometry, structure and wave propagation characteristics. Loop antennas [16,17] are simple circular or planar loops that generate helicon waves by inducing azimuthal currents. They are simple to construct, but not efficient at exciting helicon modes of higher azimuthal order ($|m| > 0$) [10]. Helical antennas [12,18] are coiled structures with a defined pitch and diameter, optimised to match the helical nature of the helicon wave. These antennas are effective at exciting helicon modes and are often used in cylindrical plasmas. The birdcage antenna [19,20] is a cylindrical, symmetrical RF antenna that is commonly used in magnetic resonance imaging (MRI) applications [21]. It consists of evenly spaced conductive rods connected by end rings at both ends, resembling a birdcage. The birdcage antenna is valued for its ability to produce highly uniform fields, its tunability to the RF feeding line since it works at antenna resonance conditions [22], and its high efficiency in exciting helicon modes [23–25]. The Nagoya type-III antenna [7,12,26], the antenna considered in this work, is an antenna widely used for helicon plasma sources and has a simple cylindrical geometry with azimuthally aligned rods (Figure 1). It has been selected in this work for its simplicity of implementation compared to other helicon antennas [18,19,23] and for the good experimental results obtained in ionising gases [5–7,12]. In this study, the antenna is fed with a radiofrequency (RF) source at a frequency of 13.56 MHz, with a power level of 1 kW. Inside the cylindrical ionization chamber, argon gas is considered almost completely ionized and the focus is on the steady state regime of the antenna. The plasma is considered to be spatially uniform in order to simplify the analysis, but also because an experimental helicon discharge does not develop strong plasma inhomogeneities, but rather a fairly uniform plasma [11,12], thus this simplification is not far from a realistic situation. A uniform magnetostatic field, generated by external coils, is directed along the axial direction, enabling the propagation of helicon waves in the plasma column. We use full-wave simulations to evaluate the real part of the antenna impedance, i.e., the antenna resistance, which is proportional to the RF power coupled to the plasma. Moreover, in the following design methods, we consider a simple cylindrical geometry with quartz tube radius *a*, antenna radius *b* and antenna length $L_a$. We also assume that, for the sake of simplicity, the plasma radius is approximately equal to the quartz tube radius (*a*).

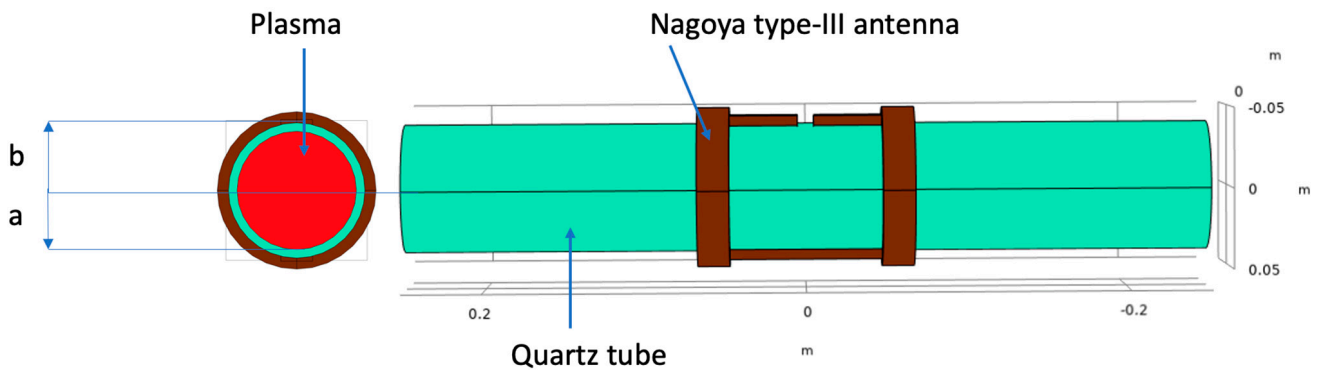

**Figure 1.** An image of the Nagoya type-III antenna (the brown cylindrical structure) wrapped around the quartz tube ionization chamber (shown in green) containing the plasma (shown in red).

*2.1. Design Method 1*

This first design method is based on the generalized theory of helicon waves [9,10]. This antenna design method takes into account the electron inertia and it is based on the evaluation of the real part of the antenna–plasma impedance [10], which is a measure of the antenna's ability to couple power into the plasma [27]. For a given discharge quartz tube and fixed plasma parameters, the deposited power into the plasma can be maximized for a given value of the antenna length, allowing the identification of an optimal antenna length, $L_{optimum}$. We assume here that the plasma is a single uniform fluid of electrons which can collide with a static background of ions and with a gas of neutrals (argon partially ionized with a single ionization charge) and the plasma is in a steady state harmonic regime, with waves represented as $\exp(i(m\theta + kz - \omega t))$, propagating in a collisional plasma with a magnetostatic field $B = B_0\hat{z}$. A cylindrical coordinate system $(r,\theta,z)$ is used and the field perturbations have cylindrical symmetry. Specifically, the helicon wave model is based on Maxwell's equations and on Ohm's law, as in the following system of equations [9,10]:

$$\begin{cases} \nabla \cdot B = 0 \\ \nabla \times E = i\omega B \\ \nabla \times B = \mu_0(J - i\omega\epsilon_0 E) = -i\omega\epsilon_0 \epsilon E \\ -i\omega m_e v = -e(E + v \times B0) - m_e v\nu \end{cases} \quad (1)$$

where E and B are the wave electric field and the wave magnetic field, respectively, $e$ is the electron charge, $\epsilon_0$ is the vacuum permittivity, $\mu_0$ is the vacuum permeability, $\epsilon$ is the complex permittivity tensor, $J$ is the current plasma density, $\omega$ is the feeding angular frequency, $m_e$ is the electron mass, v is the electron velocity and $\nu$ is the total collisional frequency. For the collisional frequency, both the collisions between electrons and ions and electrons and neutrals are considered, with the following collisional frequency expressions [28]:

$$\nu_{ei} = 3.9 \cdot 10^{-6} n_i Z_i^2 \frac{\ln \Lambda}{T_{e0}^{\frac{3}{2}}} \quad (2)$$

$$\nu_{en} = \sigma_n n_n v_{th,e} \quad (3)$$

where $Z_i$ = 1 is the ionization charge number, $ln\Lambda$ is the Coulomb logarithm that is an impact parameter defined in a collisional plasma where particles interact due to the Coulomb force, $\sigma_n$ is an atom cross-section computed as the cross-section of the neutral atom assumed spherical ($\pi r_n^2$) and measured in $m^2$, $n_n$ is the neutral density, $n_i$ is the ion plasma density, and $v_{th,e} = \sqrt{2T_{e0}/m_e}$ is the electron thermal velocity in the ionization chamber with $T_{e0}$ that is the electron plasma temperature.

Solving the system of Equation (1), one obtains a second-order vector differential equation, as follows:

$$\delta \nabla \times \nabla \times B - k\nabla \times B + k_w^2 B = 0 \quad (4)$$

where k is the parallel wave number, $k_w$ is the whistler wave number defined as $k_w = k_s \sqrt{\delta}$ with $k_s = \omega_p/c$ ("skin number"), with c the speed of light, and the $\delta$ parameter is defined as follows:

$$\delta = \frac{\omega + i\nu}{\omega_c} \quad (5)$$

where $\omega$ is the feeding angular frequency, $\nu$ is the collisional frequency and $\omega_c$ is the electron cyclotron angular frequency ($\omega_c = 2\pi f_c$ where $f_c$ is the electron cyclotron frequency). Assuming that the solution of this differential Equation (4) is such that the rotor of the wave magnetic field is parallel to the magnetic field itself by means of the total wave number $\beta$, that is $\nabla \times B = \beta B$, the differential Equation (4) can be factorized as follows, resulting in a second-order algebraic equation:

$$(\beta_1 - \nabla \times)(\beta_2 - \nabla \times)B = 0 \quad (6)$$

$$\delta\beta^2 - k\beta + k_w^2 = 0 \quad (7)$$

A particular solution of this second-order algebraic equation is obtained if the electron mass is neglected ($m_e$ = 0), which implies that $\delta$ = 0 and the solution is a single complex $\beta$ that is the helicon wave number for a collisional uniform plasma:

$$\beta = \frac{(\omega + i\nu)}{k} \frac{n_{e0} e \mu_0}{B_0} \quad (8)$$

where $n_{e0}$ is the electron plasma density. If no additional hypotheses are made, the second-order algebraic Equation (7) has two different roots, as follows:

$$\beta_{1,2} = \frac{k \mp \sqrt{k^2 - 4\delta k_w^2}}{2\delta} \tag{9}$$

These roots are well-separated if $k^2 \gg \delta k_w^2$, a hypothesis that allows us to consider the Taylor series expansion truncated at the first order of the polynomial $(1 + x)^\alpha = 1 + \alpha x$, which can be approximated as follows:

$$\beta_1 = \frac{k_w^2}{k} \qquad \beta_2 = \frac{k}{\delta} \tag{10}$$

The first solution is the helicon wave number, obtained also when the electron mass is neglected, while the other is a strongly damped cyclotron wave (i.e., the TG wave). These two waves are both propagating in the plasma when the electron inertia is retained in the model and they are coupled at the boundary (for a constant density profile) or in the plasma region (for a radial varying density profile). The relative amplitudes of the waves can be computed considering the boundary conditions for both waves. The TG waves have typically lower amplitude in the plasma core, but their amplitude is higher at the plasma edge where they strongly contribute to the wave energy absorption into the plasma. The previous second order algebraic Equation (7) for waves propagating in a magnetized collisional plasma can be also rewritten as follows:

$$k = \frac{\delta}{\beta}\left(\beta^2 + k_s^2\right) \tag{11}$$

From this Equation (11), and considering that k is a real number as done in [9,10], a maximum value $k_{max}$ for the parallel wave number k is obtained assuming the parallel wave number k is equal to the total wave number $\beta$ and also a minimum value $k_{min}$ is obtained differentiating the previous relation with respect to the total wave number $\beta$, and they are as follows:

$$k_{min} = 2\delta k_s \qquad k_{max} = \beta \tag{12}$$

This interval for k is considered for the computation of the optimal antenna length.

According to [10], the antenna–plasma resistance, a measure of how much power is transferred from the antenna to the plasma, is computed as follows:

$$R_A(m, L_a) = \int_{-\infty}^{+\infty} P_k(k, m, L_a) dk \tag{13}$$

where $P_k$ is the spectral antenna–plasma resistance:

$$P_k(k, m, L_a) = S_k(k, m) p_A(k, m, L_a) \tag{14}$$

where $S_k$ is the specific power density (independent from the antenna) and $p_A$ is the antenna power density, $m$ is the azimuthal mode number of the helicon wave and $L_a$ is the antenna length. The power transferred from the antenna to the plasma is computed as the result of the Joule heating of the plasma, according to the following formula [10]:

$$P = \int \frac{1}{2} E^* J_{plasma} dV = \frac{|I_0|^2}{2}(R_A + iX_A) \tag{15}$$

where $R_A$ and $X_A$ are the antenna–plasma resistance and reactance, respectively. The simplifying assumption made here with respect to [10] is that the leading terms in the expression of $S_k$ are those closely related to the wave dispersion, i.e., $S_k$, as a function of k, and they can be approximated with the following expression, unless scaling factors and slowly varying terms with k:

$$S_k(k,m) \approx \left| \frac{p_m(a,b,c,k)}{D(a,c,k)} \right|^2 \tag{16}$$

where *a* is the plasma radius (assumed here to be equal to the quartz tube radius), *b* is the antenna radius and *c* is the external conductive wall radius (taken as equal to ten times the plasma radius in order to neglect the influence of the wall [10]), $p_m$ is a geometric function (independent from the plasma properties) and *D* is the plasma dispersion function (independent from the antenna properties) and they have the following expressions:

$$p_m(r) = \frac{K'_m(Tr)I'_m(Tc) - K'_m(Tc)I'_m(Tr)}{K_m(Ta)I'_m(Tc) - K'_m(Tc)I_m(Ta)} \tag{17}$$

$$D(a,c,k) = F_1 G_2 - F_2 G_1 \tag{18}$$

with $F_{1,2}$ and $G_{1,2}$ that are defined as in [10]. The first factor $S_k$ in Equation (14) is the only function of the parallel wave number *k*, having here selected the propagation mode *m* = 1 that is the expected propagating helicon mode according to measurements [9,10,12]. The second factor $p_A$ in Equation (14) depends on the $kL_a$ product, being computed from the Fourier transform of the antenna current density ($K_\phi(k, m, L_a)$) according to the following expression [10]:

$$p_A(k,m,L_a) = \frac{1}{I_0^2} |K_\phi(k,m,L_a)|^2 \tag{19}$$

where $I_0$ is the antenna current. $K_\phi(k, m, L_a)$ depends on the antenna geometry and on the azimuthal mode number considered [10]. For a Nagoya type-III antenna with zero turns ($\theta$ = 0) and propagation mode *m* = 1, the Fourier transform of the current density has the following expression:

$$K_\phi(k,m,L_a) = -\frac{2}{\pi} I_0 \sin\left(\frac{kL_a}{2}\right) \tag{20}$$

This is a periodic function that is maximized for a periodic value of the product $kL_a = \pi(2h + 1)$, with *h* being an integer value (with h = 0 for the minimum length of the antenna). Thus, the first optimal length of the antenna is computed maximizing the antenna–plasma resistance (Equation (13)), where in the integration, the parallel wave number *k* is varied from its minimum to its maximum value (Equation (12)).

*2.2. Design Method 2*

A second design method, taken for comparison, is the one considered in one of the first studies on the Nagoya type-III antenna [7]. It assumes the Landau damping hypothesis, according to which the wave phase velocity is almost equal to the velocity of primary electrons contained in the neutral gas and, being at a slightly slower speed, it is possible to have an electron acceleration which leads to the ionization of the gas by electron–ion collisions. Although the Landau damping physics has no experimental confirmation in helicon discharges, we consider this design method because it was the first and best-known design method published in the literature for the Nagoya Type-III antenna [7] and also because it is still used in the literature [29]. We assume also here that the plasma is a single uniform fluid of electrons, but here we neglect the electron inertia. The argon gas is considered fully ionized, ions are at rest, the plasma is in a steady state harmonic regime with waves represented as $exp(i(m\theta + kz - \omega t))$ and the bias magnetic field is in the axial direction ($B = B_0 \hat{z}$). We also assume here that the antenna radius is equal to the quartz tube radius (*b* = *a*).

The main assumptions used in this design method are the following: the parallel phase velocity $\omega/k$ equals the thermal velocity of the primary electrons, the helicon dispersion relation for a uniform non-collisional plasma is $\alpha = (\omega/k)(\mu_0 e n_{e0}/B_0)$ and the electron energy $E_e$, measured in eV, of primary electrons interacting with the helicon wave is constrained by the dispersion relation, assuming given $B_0$, $n_{e0}$ and $a$.

The first assumption implies that the phase velocity is a function of the electron energy $E_e$, according to the following relation [7]:

$$\frac{\omega}{k} = 5.93 \cdot 10^5 E_e^{\frac{1}{2}} \qquad (21)$$

Moreover, in the dispersion relation it is assumed that the total wave number $\alpha$ is almost equal to the transversal wave number $T = 3.83/a$ (for mode m = 1) that is determined from the boundary condition $B_r(r = a) = 0$, leading to the following expression for the dispersion relation [7]:

$$\frac{B_0}{n_{e0}} = 31.2 \cdot 10^{-21} E_e^{\frac{1}{2}} a \qquad (22)$$

Finally, the antenna length is computed assuming that is equal to $\lambda/2$, where $\lambda$ is the plasma longitudinal helicon wavelength, and it is given by the following relation:

$$L_a = \frac{\pi}{k} = \frac{5.93 \cdot 10^5 \pi E_e^{\frac{1}{2}}}{\omega} \qquad (23)$$

where the energy of primary electrons $E_e$ must satisfy Equation (21). This design method, when the electron energy $E_e$ is considered as input, has the drawback of assuming an arbitrary value for this energy. Unlike the original paper [7], where the electron energy was taken as close to the peak energy of the ionization cross section of the gas, here we derived this energy from the helicon wave dispersion relation (Equation (21)), having fixed the plasma parameters and the antenna geometry, at the expense of losing the dependence of the design method on the gas properties. Moreover, from an experimental point of view, this method has less foundation, since the existence of accelerated primary electrons via Landau damping has not been confirmed experimentally [8].

*2.3. Design Method 3*

A third method is also considered for comparison. We assume also here that the plasma is a single uniform fluid of electrons and we neglect the electron inertia. The argon gas is fully ionized, ions are at rest, the plasma is in a steady state harmonic regime with waves represented as $exp(i(m\theta + kz - \omega t))$ and the bias magnetic field is in the axial direction ($B = B_0 \hat{z}$). We also assume here that the antenna radius is equal to the quartz tube radius ($b = a$). The boundary condition $B_r(Ta) = 0$, valid for both a conductive and an insulating wall, implies the computation of the first zero of the Bessel function of first kind ($J_1(Ta) = 0$ for mode m = 1), and it provides a geometrical relation between the quartz tube radius or plasma radius ($a$) and the transverse wave number ($T$), as follows:

$$T = \frac{3.83}{a} \qquad (24)$$

Therefore, when the quartz tube radius $a$ is given, the transverse wave number is fixed.

The antenna length is defined by imposing a periodic perturbation on a cylindrical plasma with the same wavelength of the plasma mode to be excited. According to the antenna mode coupling explanation given by Chen [7], the Nagoya type-III antenna is able to set up in the plasma a strong transverse electric field with a phase inversion between the two antenna end rings, enabling coupling with a plasma mode with the same wavelength. Thus, in this physical picture, we impose the following general condition for a forced periodic oscillation in the plasma:

$$L_a = \frac{(2n+1)}{2}\lambda, \; n = 0, 1, 2, 3 \qquad (25)$$

where $\lambda$ is the longitudinal helicon wavelength. Considering the quadrature relation for the total wave number ($\alpha$) and the helicon dispersion relation for a uniform plasma, the following system of equations can be obtained:

$$\begin{cases} \alpha^2 = T^2 + k^2 \\ \alpha = \frac{\omega}{k} \frac{\mu_0 e n_{e0}}{B_0} = \frac{\omega}{k} \frac{\omega_p^2}{\omega_c c^2} \end{cases} \tag{26}$$

The resultant equation of this system of equations is a biquadratic algebraic equation of the fourth order in $\alpha$:

$$\alpha^4 - T^2 \alpha^2 - \frac{\omega^2 \omega_p^4}{\omega_c^2 c^4} = 0 \tag{27}$$

This equation can be solved analytically with the following formula:

$$\alpha = \pm \sqrt{\frac{T^2 + \sqrt{T^4 + 4\frac{\omega^2 \omega_p^4}{\omega_c^2 c^4}}}{2}} \tag{28}$$

Now, considering the dispersion relation for the helicon wave in a noncollisional plasma that can be rewritten as follows:

$$k = \frac{\omega}{\alpha} \frac{\omega_p^2}{\omega_c c^2} \tag{29}$$

one can compute, from the general condition (Equation (25)) with n = 0 (minimum length of the antenna), the antenna length as a function of the plasma longitudinal helicon wavelength:

$$L_a = \frac{\lambda}{2} = \frac{\pi}{k} \tag{30}$$

where we considered $\lambda = 2\pi/k$. Thus, this design method allows us to define the cylindrical geometry of the Nagoya type-III antenna using as input the plasma state ($n_{e0}$,$B_0$) in which the helicon waves are supposed to be, the feeding angular frequency ($\omega$) and the quartz tube radius (*a*).

## 3. Simulations and Results

The three previous sizing methods are compared with numerical results from full-wave 3D electromagnetic simulations made with a finite element method (FEM) solver based on the Partial Differential Equation Toolbox of MATLAB [30]. This solver is used to solve Maxwell's equations in an equivalent dielectric medium for a 3D geometry, allowing us to simulate the electromagnetic antenna coupling and propagation of helicon waves in the magnetized plasma [31]. In these simulations, the plasma is modelled as a uniform conductive anisotropic medium whose permittivity is complex and assigned in tensor form, according to the cold-plasma dielectric tensor provided from Stix [32] with collisional corrections [27,33], which has the following expression:

$$\epsilon = \begin{bmatrix} S & jD & 0 \\ -jD & S & 0 \\ 0 & 0 & P \end{bmatrix} \tag{31}$$

where the analytical expressions for the tensor elements S, D and P are as follows:

$$\begin{cases} S = 1 - \sum_{a=e,i} \frac{\omega_{p,\alpha}^2 (\omega - j\nu_\alpha)}{\omega \left[ (\omega - j\nu_\alpha)^2 - \omega_{c,\alpha}^2 \right]} \\ D = \sum_{a=e,i} \frac{\sigma_\alpha \omega_{c,\alpha}}{\omega} \frac{\omega_{p,\alpha}^2}{(\omega - j\nu_\alpha)^2 - \omega_{c,\alpha}^2} \\ P = 1 - \sum_{a=e,i} \frac{\omega_{p,\alpha}^2}{\omega(\omega - j\nu_\alpha)} \end{cases} \tag{32}$$

where the subscript $\alpha$ refers to the species forming the plasma, $\omega_{p,\alpha}$ is the plasma frequency, $\omega_{c,\alpha}$ is the cyclotron angular frequency, $\sigma_\alpha = \pm 1$ is the particle charge sign and $\nu_\alpha$ is the collision frequency. The numerical code has been validated by comparing the obtained results with those from a previous study [27]. In this validation phase, the Nagoya type-III antenna geometry and plasma parameters were chosen as in [27], with inner quartz tube radius a = 2 cm, external quartz tube radius b = 3 cm (equal to the antenna radius), and tube thickness t = 1 cm. The antenna length was assumed to be equal to $L_a$ = 5 cm, the plasma is confined with a magnetostatic field equal to $B_0$ = 100 mT with an electron temperature $T_{e0}$ = 3 eV, a neutral pressure $p_n$ = 2 Pa and a neutrals temperature of $T_n$ = 298 K. The feeding RF power is provided at a frequency equal to $f$ = 15 MHz with a feeding voltage of $V$ = 1 V in ideal impedance-matching conditions (as in [27]). The obtained results are in good agreement with [27], as shown in Figure 2.

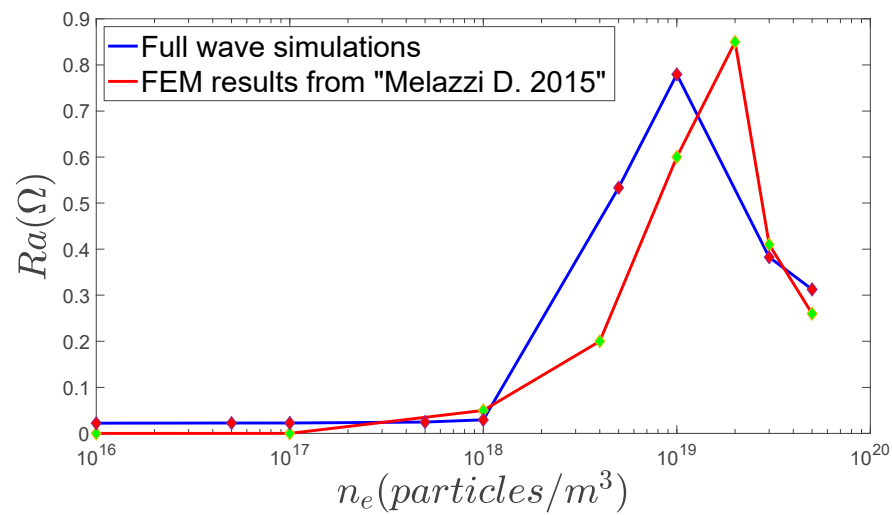

**Figure 2.** Comparison of antenna resistance values for a Nagoya type-III geometry, for validation of the FEM code used in this work. ($B_0$ = 100 mT, $p_n$ = 2 Pa, b = 3 cm, $L_a$ = 5 cm, $T_e$ = 3 eV, $f'$ = 15 MHz [27]).

Starting from this 3D simulation model, a parametric full-wave simulation of the antenna length ($L_a$) has been performed in order to evaluate the real part of the antenna impedance. We consider three simulation test cases. The first case is the reference case with typical helicon discharge parameters ($n_{e0}$ = 1.0 × 10$^{18}$ particles/m$^3$, $B_0$ = 10 mT), called "Case 1". The second case is a simulation with a higher plasma density (5 × 10$^{18}$ particles/m$^3$) with respect to the reference case, called "Case 2". The third case, called "Case 3", is a simulation with a larger magnetic field (15 mT) with respect to the reference case. The parameters considered for the design of the Nagoya type-III antenna in the three simulation test cases are reported in Table 1. The simulation results for the three simulation test cases are reported in Figure 3. These plots show a clear peak of the antenna resistance corresponding to a first optimal length of the Nagoya type-III antenna.

For each simulation test case, the optimal antenna length from full-wave simulations is compared with those computed with the aforementioned three design methods and the corresponding values are reported with vertical lines in Figure 3. In Table 2 are reported all comparisons among antenna lengths, reporting the relative error (as a percentage) with respect to the reference antenna length from full-wave simulations and also the average relative error over all the three test cases for each method. Relative to Design Method 1, in Figure 4, the contour plot of the $S_k$ function in the Gauss plane is reported, showing several peaks of the $S_k$ function, each corresponding to different helicon propagating modes. Thus, it is possible to observe that more than one propagating helicon mode is allowed for each simulation case. Design Method 1 seems the more accurate in the evaluation of the optimum length; in fact it has the lowest average relative error over all the three test cases. Moreover, we observe that in the second simulation test case, the simpler Design Method 3 is capable of predicting the optimal length with the least error. This good prediction capability of the last method is related to the high electron density of "Case 2" for which it is known [9,10] that the simple uniform plasma model, without taking into account the coupling with the TG wave, is a quite accurate description for the helicon waves.

**Table 1.** Design parameters.

|  | **Case 1** | **Case 2** | **Case 3** |
|---|---|---|---|
| Species | Argon | Argon | Argon |
| f | 13.56 MHz | 13.56 MHz | 13.56 MHz |
| $B_0$ | 10 mT | 10 mT | 15 mT |
| $n_{e0}$ | $1.0 \times 10^{18} \frac{\text{particles}}{\text{m}^3}$ | $5.00 \times 10^{18} \frac{\text{particles}}{\text{m}^3}$ | $1.00 \times 10^{18} \frac{\text{particles}}{\text{m}^3}$ |
| $T_{e0}$ | 3 eV | 3 eV | 3 eV |
| $p_n$ | 2 Pa | 2 Pa | 2 Pa |
| b | 4.5 cm (antenna, Figure 1) | 4.5 cm (antenna, Figure 1) | 4.5 cm (antenna, Figure 1) |
| a | 4.0 cm (quartz tube, Figure 1) | 4.0 cm (quartz tube, Figure 1) | 4.0 cm (quartz tube, Figure 1) |

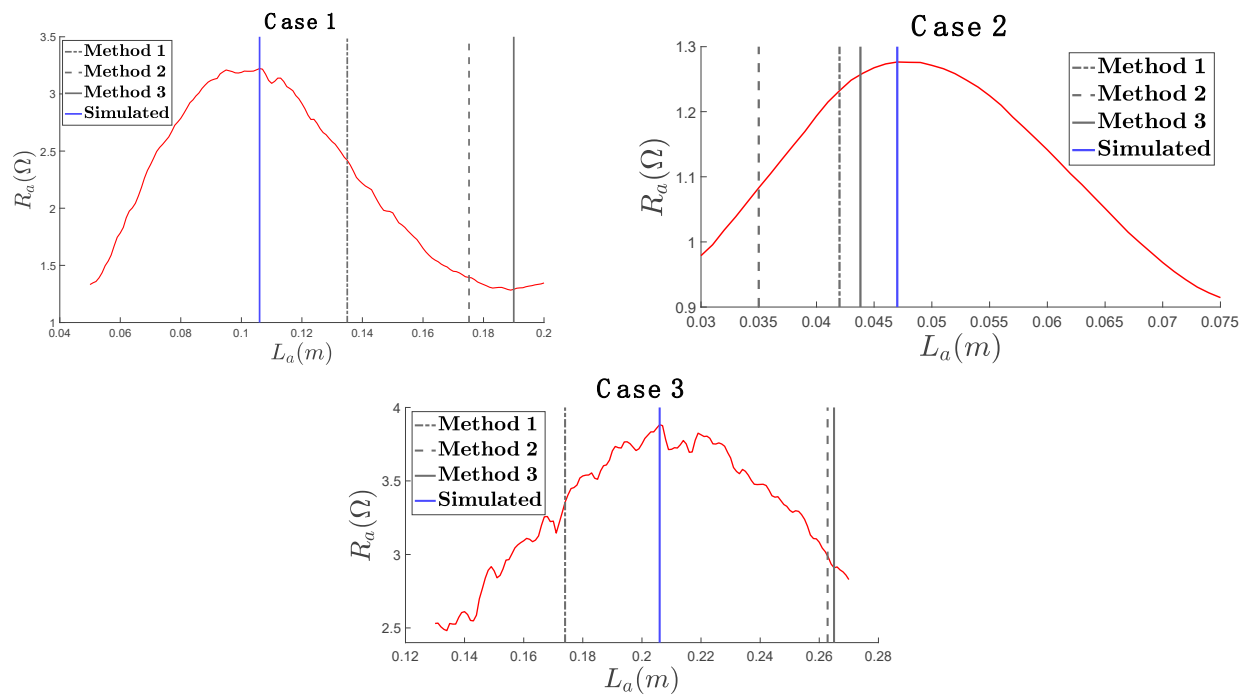

**Figure 3.** The antenna resistance ($R_a$, shown in red) computed as a function of the antenna length ($L_a$) for the three simulation test cases.

**Table 2.** Comparison of the antenna length evaluation for the three design methods and the simulations.

| - | **Case 1** | **Case 2** | **Case 3** | **Average Error** |
|---|---|---|---|---|
| Method 1 | 13.50 cm (27.36%) | 4.20 cm (−12.5%) | 17.40 cm (−15.53%) | 18.46% |
| Method 2 | 17.52 cm (65.28%) | 3.50 cm (−27.08%) | 26.28 cm (27.57%) | 39.98% |
| Method 3 | 19.00 cm (79.25%) | 4.38 cm (−8.75%) | 26.50 cm (28.60%) | 38.87% |
| Simulated | 10.60 cm | 4.80 cm | 20.60 cm | - |

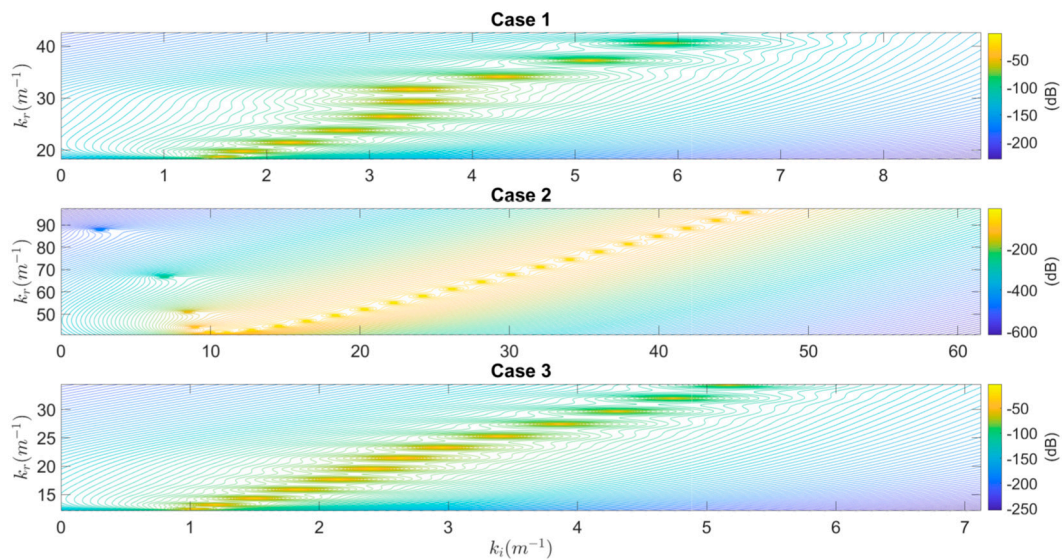

**Figure 4.** Modulus (in dB) of the function $S_k$ in the Gauss plane, for the three simulation test cases.

## 4. Conclusions

In this work, taking into account previous studies on helicon wave theory [7,9,10,12], we described three different design methods for the Nagoya type-III antenna, each based on different simplifying modelling assumptions, and we compared the antenna length computed with these methods with the optimal antenna length computed with full-wave 3D simulations for three different plasma configurations. In particular, we observe that Design Method 1, which takes into account the wave coupling among helicon waves and TG waves, seems the more accurate in the evaluation of the optimal length; in fact, it has the lowest average relative error over all the three test cases. However, we must also observe that Design Method 3, which takes into account only the propagation of the helicon wave, is quite accurate for the high-density plasma case, while Design Method 2, which uses the Landau damping hypothesis, is the least accurate. Thus, we must conclude that, for quite common plasma parameters of helicon discharges, it is possible to use Design Method 1 with good accuracy for a first sizing of the Nagoya type-III antenna, while, for higher density plasmas, it is also possible to obtain a quite accurate first sizing of the Nagoya type-III antenna with the simpler Design Method 3. In relation to the experimental results reported in [19], where it was found that the parallel wavelength of the excited helicon modes does not seem to be determined by the antenna length, but instead is strongly determined by the plasma density and the magnetic field, we observe that the more accurate Design Method 1 seems to be compatible with such findings since it shows clearly how a fixed antenna length can excite a full spectrum of propagating helicon modes in the plasma, as shown in Figure 4, with different parallel wavelengths for different plasma parameters, as was observed in [19]. However, Design Method 1 shows analytically (this is also confirmed numerically by full-wave simulations) that, in accordance with previous theoretical [10] and experimental [6,12,18] results, there is always room for the definition of an optimal antenna length, which in turn must be more or less suboptimal when working with plasma densities and magnetic fields that deviate to different degrees from the design plasma parameters. In conclusion, the aim of this work was to provide a practical and fast method for dimensioning the antenna length of the Nagoya type-III antenna, and we found that Design Method 1 is a good candidate for the sizing of the Nagoya type-III antenna. Design Method 1 can thus be useful for initializing more accurate but computationally heavier full-wave simulations in 3D geometry or simply for a rapid prototyping of the helicon antenna. Moreover, these numerical results can be useful for the development of a helicon thruster but also for the design of a high-density RF plasma source, when both applications are employing a Nagoya type-III antenna. In a future study, we will extend this first analysis to different types of helicon antennas, performing also a more extensive analysis of Design Method 1 over a wider plasma parameter space.

**Author Contributions:** Conceptualization, D.I. and F.N.; Methodology, A.C., D.I. and F.N.; Investigation, D.I. and F.N.; Writing—original draft, D.I. and F.N.; Writing—review & editing, A.C., A.D.N., A.I., D.I., F.N. and S.M. All authors have read and agreed to the published version of the manuscript.

**Funding:** This research received no external funding.

**Data Availability Statement:** The original contributions presented in this study are included in the article. Further inquiries can be directed to the corresponding author.

**Conflicts of Interest:** The authors declare no conflict of interest.